\documentclass[aps,preprint,showpacs, showkeys]{revtex4}%
\usepackage{amsfonts}
\usepackage{amsmath}
\usepackage{amssymb}
\usepackage{graphicx}
\usepackage{epsfig}
\usepackage{exscale}
\usepackage{float}
\usepackage{bbm}%
\providecommand{\U}[1]{\protect\rule{.1in}{.1in}}

\begin{document}
\title[ ]{ Effective Action study of $\mathcal{PT}$-Symmetry Breaking for the
non-Hermitian $\left(  i\phi^{3}\right)  _{6-\epsilon}$ Theory and The
Yang-Lee Edge Singularity}
\author{Abouzeid M. Shalaby}
\email{amshalab@qu.edu.qa}
\affiliation{Department of Mathematics, Statistics, and Physics, Qatar University, Al
Tarfa, Doha 2713, Qatar}

\begin{abstract}
We use the effective potential method to study the $\mathcal{PT}$-symmetry
breaking of the non-Hermitian $i\phi^{3}$ field theory in $6-\epsilon$
space-time dimensions. The critical exponents so obtained coincide with the
exact values listed in the literature. We showed that at the point of
$\mathcal{PT}$-symmetry breaking, the vacuum-vacuum amplitude is certainly
zero and the fugacity is one which mimics a Yang-Lee edge singularity in
magnetic systems. What makes this work interesting is that it takes into
account problems which are always overlooked in the literature for the
Yang-Lee model like stability, unitarity and generation of Stokes wedges at
space-time dimensions for which divergences occur in the theory . Besides,
here we make direct calculation of critical exponents from the dependance of
the order parameter on external magnetic field not from the density of zeros
of the partition function.

\end{abstract}
\maketitle

The paper of Carl Bender and Stefan Boettcher in Ref.\cite{bendr} sparked what
we can call an actual start of the possible acceptance of non-Hermitian
theories to play a role in nature's description. This paper stimulated the
interest of many researchers to study the $\mathcal{PT}$-symmetric theories
which led to the growing of their applications in different branches in
Physics. In fact, the importance of employing such theories to play a role in
nature's description might go back to Dirac \cite{Dirac} who tried to
introduce a finite quantum field theory via the introduction of auxiliary
fields that turn the theory non-self adjoint. Then Pauli followed Dirac with
more details \cite{Pauli}. Lee and wick then introduced a finite QED
\cite{lee1, lee2} via the inclusion of abnormal fields (Lee-Wick fields) which
besides of turning the theory non-self adjoint, it makes the theory possessing
ghost states. In 2007, the idea has been extended to the standard model via an
extension called the Lee-Wick standard model \cite{lee-wick} which solves the
Hierarchy problem but suffers from the existence of ghost states too. In
Ref.\cite{ghost}, we showed that one can use the $\mathcal{PT}$-symmetric
tools to overcome the ghost states problem in such theories. In the Lee-Wick
standard model however, they add higher derivative terms to the Lagrangian
instead of adding Lee-Wick field directly but it has been shown that the
Lagrangian can be rewritten in terms of a normal and a Lee-Wick field which in
turn shows that the theory is not self-adjoint. Moreover, recently the idea
has been extended to gravity and a super-renormailzable gravity has been
obtained via the insertion of higher derivative terms \cite{gravity}. So, it
seems that the importance of studying $\mathcal{PT}$-symmetric theories goes
beyond considering it as a mathematical training and might bear the solution
of existing problems appearing in models that are constrained to Hermitian
theories only. However, in space-time dimensions higher than one (quantum
field theory), the tools followed to study $\mathcal{PT}$-symmetric quantum
problems might not work. For instance, if one generates the Stokes wedges for
a theory in one dimension (Quantum mechanics) through the constraint of finite
controlling integral \cite{Klav}, the quantum field version of the same theory
will not have finite controlling integral due to existence of divergences in
the theory. Accordingly, one needs to test a technique that can mimic the
selection of a contour in the complex field plane within the Stokes wedges as
well as being able to remedy the divergences in the theory. The suggested tool
we follow here is the effective action formulation and shall try to test its
predictions via comparison with indirect experimental results. The model we
select to study in this work is the $\mathcal{PT}$-symmetric $i\phi^{3}$
theory. In $0+1$ space-time dimensions, one can find recent rigorous work in
the literature that stressing that model
\cite{crossing,Dorey,Dorey2,Shin1,Shin2,pade,spect,Benderx3,Zadahx3}. In
Higher space-time dimensions, one always confronted by mainly two problems in
studding a $\mathcal{PT}$-symmetric field theory. The first one is the
generation of the Stokes wedges of the theory where divergences will prevent
us from having a finite integral representing the generating function. The
other problem is the calculation of the metric operator in $\mathcal{PT}%
$-symmetric field theory where it is hard to obtain. As we will see in this
work, the effective action technique seems to be the most suitable tool to
study a $\mathcal{PT}$-symmetric field theory.

The experimental test of predictions within $\mathcal{PT}$-symmetric theories
have been stressed in many articles (for instance Refs. \cite{ncj,npj,pra}).
In fact, there is a growing interest in that direction and it seems to be a
matter of time to the full acceptance of non-Hermitian theories to share in
solving existing problems in the current Hermitian description of a natural
phenomena. A theoretical test (or indirect experimental test) can be offered
by comparing critical behaviors of a $\mathcal{PT}$-symmetric \ model with
another model that lies in the same class of universality.

The bridge between $\mathcal{PT}$-symmetric theories and Ising model has been
offered by Fisher \cite{LYsing1,prdphi3} who identified the effective action
of the Magnetization by a Landau-Ginzberg theory given by \cite{Musardob}:%

\begin{equation}
S=\int dx^{d}\left(  \frac{1}{2}\left(  \partial_{\mu}\phi\right)
^{2}+i\left(  h-h_{c}\right)  \phi+ig\phi^{3}\right)  , \label{Lee-Yang-M}%
\end{equation}
where $h$ represents the external magnetic field. This effective field theory
has reproduced the critical behavior associated with the Yang-Lee edge
singularity \cite{Young-Lee1,Young-Lee2}. In fact, at the continuum limit, the
zeroth of the partition function in the fugacity complex plane can touch the
real axis and the intersection with the real axis represents a critical point.
At that limit, the zeros are more dense and the density of zeros follows a a
critical formula from which one can extract the critical exponent
$\sigma=\frac{1}{\delta}$ \footnote{See Eq. (14.5.3) and Eq. (14.5.3) in Ref.
\cite{Musardob}}.

The Yang-Lee model in Eq.(\ref{Lee-Yang-M}) has been extensively studied in
the literature regarding the edge singularity
\cite{Musardob,Cardy,Cardy2,Wipf} but most of the studies always skip
important problems like unitarity and the needed $\mathcal{PT}$-symmetric
boundary conditions. Unitarity can be solved via the insertion of the metric
operator $\eta$ such that the physical amplitudes associated with an operator
$\hat{O}$ take the form $\langle\hat{O}\rangle=\langle\psi\left\vert \eta
\hat{O}\right\vert \psi\rangle$ \cite{Zadah}. Such formulation leads to a
Hilbert space for which orthonormallity is satisfied as well as unitarity is
guaranteed. However, the metric operator is always hard to get for a quantum
field theory but it has been argued by Jones that certain techniques can be
used to obtain physical amplitudes for which the metric operator is
automatically implemented \cite{jonesgr2}. Regarding the other problem, in
order to quantize a $\mathcal{PT}$-symmetric theory one needs to subject it to
a boundary condition in the complex-field plane. The boundary condition is set
by selecting a complex contour in the complex-field plane over which the
controlling integral $\int D\phi\exp\left(  -\int dx^{d}i\phi^{3}\left(
x\right)  \right)  $ exists \cite{Klav}. In fact, it is always stated that
this integral is given by the multiplication of infinite number of the one
dimensional integral $\int_{c}d\phi_{i}\exp\left(  -i\phi^{3}\left(
x_{i}\right)  \right)  $. Now, if one selects a contour over which the low
dimensional integral $\int_{c}d\phi_{i}\exp\left(  -i\phi^{3}\left(
x_{i}\right)  \right)  $ is set finite, this does not guarantee that the full
integral $\int D\phi\exp\left(  -\int dx^{d}i\phi^{3}\left(  x\right)
\right)  $ is finite. This is a manifestation of the existence of divergences
in quantum field theories at relatively high space-time dimensions.
Accordingly claiming that the Yang-Lee model in space-time dimensions higher
than two is a real-line theory is not assured \footnote{In future, We plane to
study if there are available complex contours over which the full controlling
factor integral exists. If so, one might obtain a finite field theory via
boundary conditions in the complex field plane}. So the challenge in studding
the Yang-Lee model in $6-\epsilon$ space-time dimensions is to use an
algorithm that overcomes all of these problems and at the end can be tested
via comparison of its predictions of the critical exponents with those found
in the literature. This is the main aim of our work here.

In our study we shall use path integral formulation of quantum field theory.
The point is that the partition function in statistical models has its counter
part in quantum field theory where in path integral formulation of a quantum
field theory it is the generating functional (the vacuum persistent amplitude)
which is given by \cite{Peskin}:%
\begin{equation}
Z(J)=\int D\phi\exp\left(  i\int d^{d}x\left(  \mathcal{L}\left[  \phi\right]
+iJ\phi\right)  \right)  . \label{Partf}%
\end{equation}
In Euclidean space it takes a form $Z(J)=\exp\left(  -S_{E}\right)  $, where
$S_{E}$ is the Euclidean form of the action. The critical phenomena in a
$\mathcal{PT}$-symmetric quantum field model can be investigated by monitoring
the $\mathcal{PT}$-symmetry breaking which is expected to be associated with a
zero of $Z(J)$ or more specifically with a Yang-Lee Edge Singularity. Instead
of studying the critical phenomena via the density of zeros of the partition
function we shall study the behavior of the order parameter versus the
coupling $J$ where as we will see in this work that $\mathcal{PT}$-symmetry
breaking occurs for the $i\phi^{3}$ field theory at a zero of $Z(J)$ and exact
critical exponents are extracted from our calculations using path integral
formulation of the theory. In fact, at the continuum limit the zeros can touch
the real axis in the complex fugacity plane, a fact that will be reflected
clearly in our calculations. Another point for following the path integral
formulation to investigate the $\mathcal{PT}$-symmetry breaking of the
Yang-Lee model is that the effective action technique associated with it can
account for the selection of a complex contour in the complex field plane but
with well known methods to cure the divergences.

Before we start our calculation we give a brief introduction for
$\mathcal{PT}$-symmetric Hamiltonians. A Hamiltonian is said to be
$\mathcal{PT}$-symmetric if it is invariant under the application of both
parity $\left(  x\rightarrow-x\right)  $ and time reversal operations. A
$\mathcal{PT}$-symmetric and non-Hermitian Hamiltonian can have real spectrum
as long as the $\mathcal{PT}$-symmetry is exact. In changing parameters of
that theory one can break this symmetry and after a point of level crossing
the energy spectrum appears in complex conjugate pairs \cite{crossing,
Dorey,Dorey2}. Moreover, at the critical point the wave function is
self-orthogonal. For the purpose of quantization, such theories follows
certain boundary conditions either in the complex-x plane or the complex-field plane.

The usual recipe for the quantization of a $\mathcal{PT}$-symmetric theory is
to subject it to the boundary condition $\psi\left(  x\right)  \rightarrow0$
as $\left\vert x\right\vert \rightarrow0$ in the complex $x-$plane. Such
condition selects regions in the complex plane called Stokes wedges from which
a contour is selected over which the calculations are carried out. However,
for the study of quantum field $\mathcal{PT}$-symmetric theory, it would be
better to follow path integral formulation since it has its own boundary
condition\cite{Klav}. In such case, the Stokes wedges are generated by
demanding the existence of the integral $\int d\phi\exp\left(  -V\left(
\phi\right)  \right)  $, where $\phi$ is the scalar field while $V\left(
\phi\right)  $ is the classical potential. In other words one needs $V\left(
\phi\right)  \rightarrow\infty$ as $\left\vert \phi\right\vert \rightarrow
\infty$ \ in the complex-field plane. Without loss of generality, one selects
contours over which the classical potential is bounded from below as a
functional of $\left\vert \phi\left(  x\right)  \right\vert $. In higher
dimensions the dominant integral in the partition function is an infinite
multiplications of the integral $\int d\phi\exp\left(  -V\left(  \phi\right)
\right)  $ and it is not guaranteed that the whole functional integral is
finite for a contour selected by the necessity of the existence of the the one
dimensional integral $\int d\phi\exp\left(  -V\left(  \phi\right)  \right)  .$
\ Moreover, exact calculations in quantum field theory are always beyond
achievement and thus one resorts to perturbative or non-perturbative methods.
So instead of selecting a contour over which to make exact functional
integrations (which is hard for most of quantum field models), one can follow
the well-known effective action regime of calculations \cite{Peskin}. The
corresponding effective potential is satisfying the bounded from below
condition and thus can mimic the boundary condition mentioned above. In this
situation the complex field has a real part that depends on position while the
imaginary part is represented by the classical field (vacuum condensate). The
equations obtained by demanding the effective potential to be bounded from
below might have many solutions for the vacuum condensate which represent
different phases of the theory. In fact, making the field $\phi\left(
x\right)  $ goes to $\phi\left(  x\right)  +v$, where $v$ is the classical
field is a kind of a point canonical transformations like the selection of a
complex contour in the complex plane. In quantum field theory, canonical
transformations can be inequivalent and thus represent different phases of the
theory \cite{thermofield}. So, in this work, we apply the effective action
method to study the Yang-Lee model of the $\mathcal{PT}$-symmetric $i\phi^{3}$
field theory in $6-\epsilon$ space-time dimensions. Other reasons behind such
selection are that the effective action formalism implements the metric in its
calculations and the similarity of the vacuum persistence amplitude to the
partition function in quantum statistical systems which enables us to study
the critical behavior of the model. Up to the best of our knowldge, there is
no other technique used in literature to study the Yang-Lee model and
take-caring of the metric and respects the boundary condition used to study
$\mathcal{PT}$-symmetric theories.

To start, consider the Lagrangian density of the $\mathcal{PT}$-symmetric
$i\phi^{3}$ field given by the form:%
\begin{equation}
\mathcal{L}\left[  \phi\right]  =\frac{1}{2}\left(  \partial\phi\right)
^{2}-\frac{1}{2}m^{2}\phi^{2}(x)-\frac{ig}{3}\phi^{3}\left(  x\right)
+iJ\phi\left(  x\right)  ,
\end{equation}
where $m$ is the bare mass of the field, $g$ is the coupling constant and $J$
is a source term that resembles the external magnetic field in magnetic
systems. Here we put the source $iJ$ to make the theory $\mathcal{PT}%
$-symmetric . The vacuum persistence amplitude is given in Eq.(\ref{Partf}).
The vacuum energy $E[J]$ can be introduced via the relation $Z(J)=\exp\left(
-iE[J]\right)  $ while the effective action $\Gamma\left(  v\right)  $ can be
obtained through the Legendre transformation of the form \cite{Peskin}:%
\begin{equation}
\Gamma\left(  v\right)  =-E[J]-i\int d^{4}yJ\left(  y\right)  v\left(
y\right)  ,
\end{equation}
where $v$ is the vacuum condensate. In assuming that the vacuum is
translational invariant then the effective potential can be introduced as
$V_{eff}=-\frac{\Gamma\left(  v\right)  }{VT},$ where $VT$ is the size of the
space-time region over which the functional integral is to go over. Following
the steps in Ref.\cite{Peskin}, one can obtain the one-loop effective
potential as:%
\begin{equation}
V_{eff}\left(  v\right)  =\left[  V\left(  \phi\right)  +\frac{i}{2}\left(
i\frac{\Gamma\left(  -\frac{d}{2}\right)  }{\left(  4\pi\right)  ^{\frac{d}%
{2}}}\left(  \frac{\partial^{2}V\left(  \phi\right)  }{\partial\phi^{2}%
}\right)  ^{\frac{d}{2}}\right)  \right]  _{\phi=v},
\end{equation}
where $d$ is the dimension of the space-time and $\Gamma$ here (not to be
confused with the effective action $\Gamma\left(  v\right)  $) is the gamma
function. For the theory under consideration, $V_{eff}\left(  v\right)  $ (it
is the vacuum energy density) takes the form:%
\begin{equation}
V_{eff}\left(  v\right)  =\frac{1}{2}m^{2}v^{2}-\frac{g}{3}\left(  iv\right)
^{3}+iJv+\frac{i}{2}\left(  i\frac{\Gamma\left(  -\frac{d}{2}\right)
}{\left(  4\pi\right)  ^{\frac{d}{2}}}\left(  m^{2}+2igv\right)  ^{\frac{d}%
{2}}\right)  . \label{effp}%
\end{equation}
The effective action satisfies the condition $\frac{\delta\Gamma\left(
v\right)  }{\delta\left(  v\right)  }=0$ or equivalently the effective
potential above satisfies the relation:%
\begin{equation}
\frac{\partial V_{eff}\left(  v\right)  }{\partial v}=0.
\end{equation}
The renormalized mass $M$ is also given by:%
\begin{equation}
\frac{\partial^{2}V_{eff}\left(  v\right)  }{\partial v^{2}}=M^{2}.
\end{equation}
As long as the renormalized mass is chosen positive so these two equations
define a minimum of the effective potential and thus can be considered a
stable one. These conditions are the counter part of the boundary condition
used in $\mathcal{PT}$-symmetric quantum mechanics and the condition that the
integral $\int_{C}d\phi\exp\left(  -V\left(  \phi\right)  \right)  $ do exist
on a complex contour $C$ in the complex $\phi$ plane. Thus the effective
action formalism bears the spirit of the boundary condition in both
$\mathcal{PT}$-symmetric quantum mechanics and quantum field theory. However,
the effective action can be turned finite for a normalized $\mathcal{PT}%
$-symmetric field theory via well known procedures. This property gives the
effective potential a privilege for the study of $\mathcal{PT}$-symmetric
field theory.

Let us go back to the effective potential in Eq.(\ref{effp}). In $6-\epsilon$
space-time dimension, this effective potential is divergent and a
regularization scheme is needed followed by a renormalization procedure. To do
that we expand in powers of $\epsilon=d-6$ to get:%

\begin{align}
V_{eff}  &  \rightarrow-\frac{1}{384\pi^{3}}\frac{\left(  m^{2}+2igv\right)
^{3}}{d-6}+\frac{1}{3}igv^{3}+\frac{1}{2}m^{2}v^{2}+iJv\nonumber\\
&  -\frac{1}{768}\left(  2igv+m^{2}\right)  ^{3}\frac{\gamma-2\ln2-\ln\pi
+\ln\left(  2igv+m^{2}\right)  -\frac{11}{6}}{\pi^{3}}\allowbreak\allowbreak\\
&  +O\left(  d-6\right)  .\nonumber
\end{align}
In using the modified minimal subtraction method ($\overline{MS}$) and
introducing a renormalization mass scale $\mu$, one gets:%
\begin{equation}
V_{eff}=\frac{1}{2}m^{2}v^{2}-\frac{g}{3}\left(  iv\right)  ^{3}+iJv-\frac
{1}{768\pi^{3}}\left(  2igv+m^{2}\right)  ^{3}\ln\left(  \frac{2igv+m^{2}}%
{\mu^{2}}\right)  .
\end{equation}
In applying the condition $\frac{\partial V_{eff}}{\partial v}=0,$ we get the
relation:%
\begin{equation}
m^{2}v+igv^{2}+iJ-\frac{1}{384}\frac{i}{\pi^{3}}g\left(  3\ln\frac{1}{\mu^{2}%
}\left(  m^{2}+2igv\right)  +1\right)  \left(  m^{2}+2igv\right)  ^{2}=0,
\label{b0}%
\end{equation}
while the mass renormalization condition gives the equation:%
\begin{equation}
m^{2}+2igv+\frac{1}{192\pi^{3}}g^{2}\left(  6\ln\frac{1}{\mu^{2}}\left(
m^{2}+2igv\right)  +5\right)  \left(  m^{2}+2igv\right)  \allowbreak=\mu^{2},
\label{mus}%
\end{equation}
where with no loss of generality we set the renormalized mass $M$ to coincide
with the renormalization scale $\mu$.

A good test for our calculations can be achieved via comparison of the
critical exponents with those known in the literature. So before we try to
solve the above equations for the classical field $v$ and the renormalized
mass $\mu$ we need to remind ourselves by the critical behavior of a magnetic
(Ising model) system undergoing a phase transition. At the critical isotherm,
the magnetization $M$ has power law dependence of the external field $J$ as
$M\sim J^{\frac{1}{\delta}}$ \cite{LYsing1} and the correlation length follows
the formula $\zeta_{gap}\sim J^{-\nu_{c}}$\cite{Pal}. The free energy per unit
volume has a mass dimension $d$ and thus it has a relation of the form
$\zeta_{gap}^{-d}\sim J^{-d\nu_{c}}$ \cite{Kaku}. In Ref.\cite{LYsing1}, it
has pointed out that the Yang-Lee model is the Landau-Ginzburg approximation
that can describe the critical behavior of the ferromagnetic materials. To
check the validity of our results we will try to obtain the critical exponents
for $\mathcal{PT}$-symmetric $i\phi^{3}$. In fact, the critical isotherm is
equivalent to set the parameter $m$ to zero and then we solve Eqs.(\ref{b0}%
,\ref{mus}) to get:%
\begin{align}
v  &  =\pm i\frac{J^{\frac{1}{2}}}{\sqrt{g+\frac{1}{96}\frac{1}{\pi^{3}}%
g^{3}+\frac{1}{32\pi^{3}}g^{3}\ln\frac{g}{\frac{g^{3}}{\left(  2\pi\right)
^{3}}\operatorname{LambertW}\left(  32\frac{\pi^{3}}{g^{2}}e^{\frac{5}{6}%
}e^{32\frac{\pi^{3}}{g^{2}}}\right)  }}},\nonumber\\
\mu &  =\sqrt{iv}\sqrt{\frac{1}{16}\frac{1}{\pi^{3}}g^{3}%
\operatorname{LambertW}\left(  32\frac{\pi^{3}}{g^{2}}e^{\frac{5}{6}%
}e^{32\frac{\pi^{3}}{g^{2}}}\right)  \allowbreak}, \label{mass}%
\end{align}
where $\operatorname{LambertW}$ is the Lambert-W function satisfying $W\left(
x\right)  \exp W\left(  x\right)  =x$. Since the renormalized mass is real,
then the vacuum condensate takes the negative imaginary sign only as expected
or
\begin{equation}
v=-i\frac{J^{\frac{1}{2}}}{\sqrt{g+\frac{1}{96}\frac{1}{\pi^{3}}g^{3}+\frac
{1}{32\pi^{3}}g^{3}\ln\frac{g}{\frac{g^{3}}{\left(  2\pi\right)  ^{3}%
}\operatorname{LambertW}\left(  32\frac{\pi^{3}}{g^{2}}e^{\frac{5}{6}%
}e^{32\frac{\pi^{3}}{g^{2}}}\right)  }}}. \label{vac}%
\end{equation}
The counter part of magnetization here is the classical field $v$ and $J$
plays the role of external magnetic field. The correlation length is
represented by inverse of mass gap $\mu$ while the free energy is represented
by the effective potential which is equivalent to vacuum energy $E_{0}:$%
\begin{equation}
E_{0}=iJv-\frac{g}{3}\left(  iv\right)  ^{3}+\frac{1}{96}\frac{i}{\pi^{3}%
}g^{3}v^{3}\ln\frac{2igv}{\mu^{2}}\sim J^{\frac{3}{2}}. \label{energy}%
\end{equation}
From Eq.(\ref{vac}), one can realize that the critical exponent $\sigma
=\frac{1}{\delta}=\frac{1}{2}$ while from Eq.(\ref{mass}) it gives
$\upsilon_{c}=\frac{1}{4}$. These are exactly the critical exponents obtained
in Ref.(\cite{LYsing1}). Moreover, from the scaling of the free energy we
should have $E_{0}\sim\mu^{d}\sim J^{\frac{6}{4}}=J^{\frac{3}{2}}$ which is
exactly what Eq.(\ref{energy}) gives.

\begin{figure}[ptbh]
\begin{center}
\epsfig{file=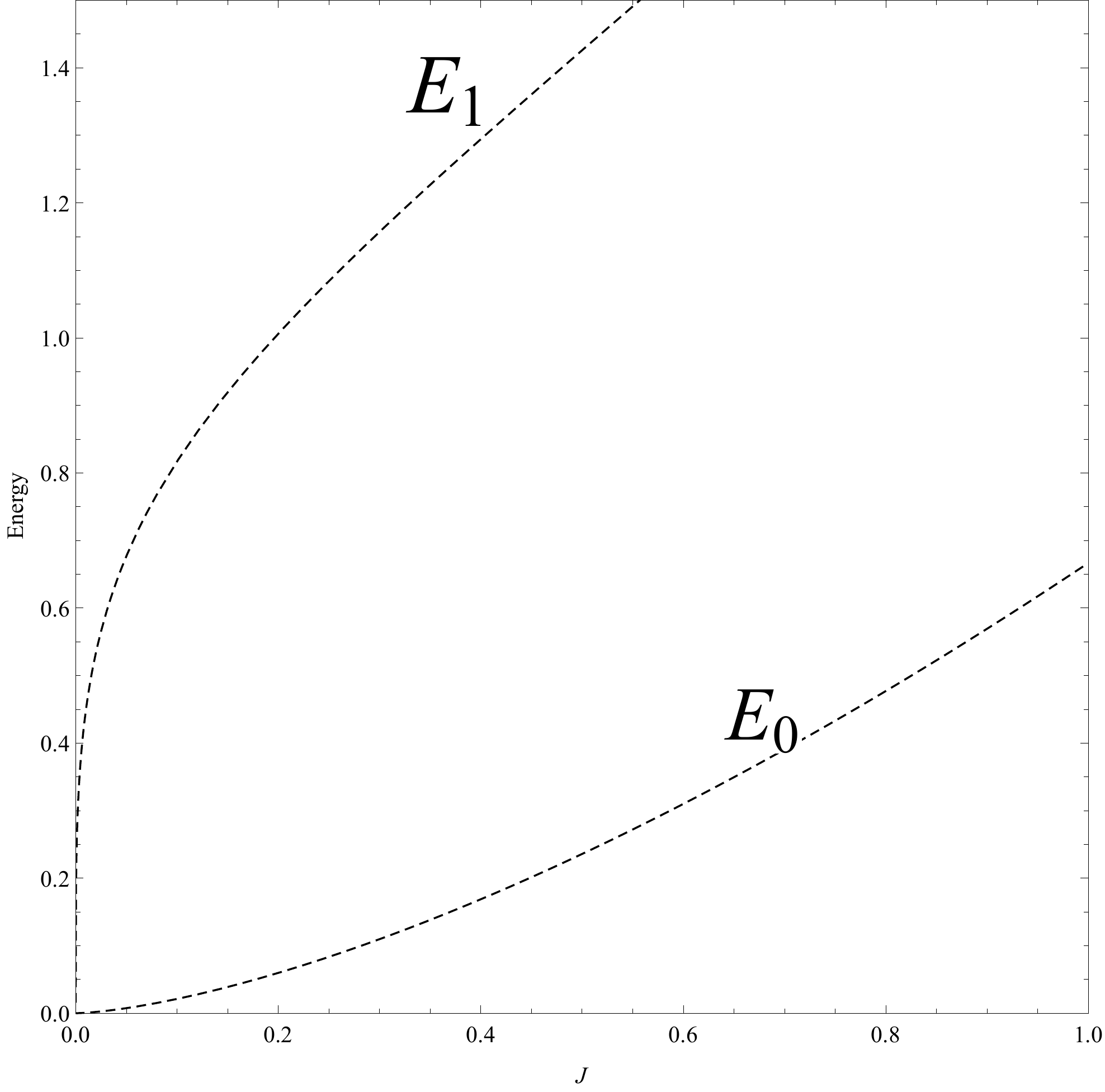,width=0.65\textwidth}
\end{center}
\caption{The first two energy levels $E_{0}$ and $E_{1}$ versus the external
magnetic field $J$ for the massless $\mathcal{PT}$-symmetric $i\phi^{3}$
theory in $6-\epsilon$ space-time dimensions.}%
\label{lcrossing}%
\end{figure}

The theory undergoes a phase transition at the point of $\mathcal{PT}%
$-symmetry breading and thus the critical point should be associated by level
crossing. To check this in our calculations, consider the lowest energy levels
$E_{0}$ and $E_{1}=E_{0}+\mu$. We plot them in Fig.\ref{lcrossing} as
functions of the external magnetic field $J$ and it is very clear that at
$J=0$, $\mathcal{PT}$-symmetry is broken. For the model under consideration
this point is supposed to be a Yang-Lee edge singularity where the zeros of
the Partition function touches the real axis of the complex fugacity. To check
that this is reflected in our calculations consider the vacuum to vacuum
amplitude given by;
\begin{equation}
Z(J)=\exp\left(  -iE[J]\right)  =\exp\left(  -iE_{0}\left(  VT\right)
\right)  .
\end{equation}
Since we renormalized the theory at a scale $\mu$ so the size of $VT$ is
$\frac{1}{i\mu^{d}}$. The factor $i$ comes from the fact that we used Wick
rotation in Feynman diagram calculation. In other words
\begin{equation}
Z(J)=\exp\left(  -\frac{E_{0}}{\mu^{6}}\right)  .
\end{equation}
This form does not explicitly depend on $J$ since from Eq.(\ref{mass}) we
obtain%
\begin{equation}
Z=\exp\left(  -\frac{256}{9}\pi^{6}\frac{3\operatorname{LambertW}\left(
32\frac{\pi^{3}}{g^{2}}\exp\left(  \frac{1}{6}\frac{5g^{2}+192\pi^{3}}{g^{2}%
}\right)  \right)  -9\ln2-1}{g^{6}\left(  \operatorname{LambertW}\left(
32\frac{\pi^{3}}{g^{2}}\exp\left(  \frac{1}{6}\frac{5g^{2}+192\pi^{3}}{g^{2}%
}\right)  \right)  \right)  ^{3}}\right)  .
\end{equation}
This form goes to zero as $g$ goes to zero too. So to check if at the critical
point ($J=0$), the partition function has a zero, one needs check that
$g\rightarrow0$ as $J\rightarrow0$. To do that we need to know the scale
behavior of the coupling $g$. In fact, we know the scale behavior of the
coupling $J$ where it goes to zero as $\mu\rightarrow0$ while to know the
scale dependance of the renormalized \ coupling $g$ we need to obtain the Beta
function of the theory. However, the Beta function can be concluded from the
one for the Hermitian $\phi^{3}$ where it has been obtained in
Ref.\cite{Collins}. Unlike the Hermitian theory, the Beta function for the
$\mathcal{PT}$-symmetric $i\phi^{3}$ is positive and up to one loop is given
by:%
\begin{equation}
\beta=\mu\frac{dg}{d\mu}=\frac{3}{512}\frac{g^{3}}{\pi^{2}}.
\end{equation}
Being positive, the $\beta-$function tells us that the theory has an infra-red
fixed point or equivalently $g\rightarrow0$ as $\mu\rightarrow0$. This
analysis assures that at the critical point ($\mu=0$, $J=0$ and $g=0$) the
partition function is zero too. Since this zero exists at $J=0$ thus the
fugacity is real (fugacity $z=1$). This shows that the critical point at which
the $\mathcal{PT}$-symmetry is broken is in fact a Yang-Lee Edge Singularity
in the sense that it resembles a zero of the partition function that touches
the real Fugacity axis as in magnetic systems. 

To conclude, We have shown that the effective action technique is the most
suitable algorithm (up to the best of our knowledge) in the literature to
study $\mathcal{PT}$-symmetric field theories specially in a relatively high
space-time dimension. The point is that for high space-time dimensions the
controlling factor used to generate the Stokes wedges of the theory in low
dimensions is no longer working. In fact for space-time dimensions for which
the theory is renormalizable we are confronted by the existence of divergences
which means that the integral defining the generating functional can not be
finite for wedges selected by the one dimensional controlling factor. So the
field is in a need to discover a technique that mimics the boundary conditions
applied to $\mathcal{PT}$-symmetric quantum mechanics or low dimensional
quantum field theory and can get rid of divergences as well. We have
illustrated that the effective potential bears these features. Then we applied
it to the normalizable Yang-Lee model in $6-\epsilon$ dimensions where we can
find predictions from studying the Yang-Lee edge singularity of the model in
the literature and thus offers a way to test the effective potential
predictions for a $\mathcal{PT}$-symmetric field theory.

To check the validity of our calculations, We obtained the critical exponents
of the $\mathcal{PT}$-symmetric $i\phi^{3}$ theory in $6-\epsilon$ space-time
dimensions and found them coincides with the exact ones listed in Ref.
\cite{LYsing1}. Moreover, we proved all the expected features associated with
$\mathcal{PT}$-symmetry breaking. For instance, we showed a level crossing at
the critical point for the first two levels of the theory. Besides, we have
shown that at the point of $\mathcal{PT}$- breaking the fugacity is real (in
fact equal one) which means that it is a Yange-Lee edge singularity where
there is a zero of the partition function that touches the real fugacity axis.

The renormalization group flow of the parameters in the theory has been
investigated. The coupling $J$ has a positive mass dimension and is expected
to go to zero as the mass scale $\mu$ goes to zero which has been reflected
clearly in our calculations. Likewise the vacuum condensate and vacuum energy
both have positive mass dimensions and have shown to go to zero as $\mu$ goes
to zero too. For the renormalized coupling $g$ on the other hand, it is
dimensionless and for its scale behavior one needs to know the $\beta
$-function of the theory. Since the $\beta$-function is positive, the theory
has an IR- fixed point or in other words $g\rightarrow0$ as $\mu\rightarrow0$
too. This fact was needed to show that the partition function $Z(J)$ goes to
zero at the critical point.

\end{document}